\documentclass{PoS}
\usepackage{amsmath}
\usepackage{amssymb}

\title{Phenomenology with a non-zero $B_s$ decay width difference}

\ShortTitle{Phenomenology with a non-zero $B_s$ decay width difference}

\author{\speaker{Robert Knegjens}\\
        Nikhef, Science Park 105, NL-1098 XG Amsterdam, Netherlands\\
        E-mail: \email{robk@nikhef.nl}}

\abstract{
The experimentally established non-zero decay width difference of the $B_s$ meson system gives us access to a mass eigenstate rate asymmetry  ${\cal A}^{f}_{\Delta\Gamma}$ for each $B_s\to f$ transition.
This observable is not only the key ingredient in converting between differing definitions of a $B_s$ branching ratio, but can also be a sensitive probe of New Physics that does not require flavour tagging.
We discuss how a pair of effective lifetimes for CP even and odd final states, which probe this asymmetry, can constrain the parameters of $B^0_s$--$\overline{B}_s^0$ mixing.
We then shift our focus to the rare decay $B_s\to \mu^+\mu^-$, for which the Standard Model branching ratio prediction receives a sizable correction due to a maximal asymmetry.
We present how this asymmetry, which can be extracted from an untagged time-dependent measurement, serves as a new observable, complementary to the branching ratio, for constraining New Physics.
Further, we analyse types of models beyond the Standard Model that this pair of $B_s\to \mu^+\mu^-$ observables can discriminate between. 
}

\FullConference{14th International Conference on B-Physics at Hadron Machines,\\
		April 8-12, 2013\\
		Bologna, Italy }

\begin{document}

% #############################################################################
\section{Introduction}

Being able to experimentally identify, or {\it tag}, the flavour states ${B}^0_s$ and $\overline{B}^0_s$ before they decay is very useful for studying their oscillations.
Unfortunately, flavour-tagging generally has a low experimental efficiency, and first requires many events to be recorded for a given $B_s \to f$ transition.
In this paper we will focus on the phenomenology that is possible before a tagged analysis becomes feasable.
To this end it is convenient to describe the $B_s$ meson system in terms of its mass eigenstates, the normal modes of this coupled system, labeled $B_{s,\rm H}$ and $B_{s,\rm L}$ for the heavier and lighter state respectively.

In the absence of a CP violating phase $\phi_s$ entering the ${B}^0_s$--$\overline{B}^0_s$ mixing amplitude, the mass eigenstates are also eigenstates of the CP operator.
This is familiar from the neutral kaon system, where this is almost the case.
In further analogy with the kaon system, 
the CP even state is similarly expected to decay more rapidly because it has access to somewhat more favourable final states.
%, such as $D_s^{(*)}\overline{D}_s^{(*)}$.
It so happens that in the Standard Model (SM) the $B_s$ mixing phase $\phi_s^{\rm SM} = (-2.1\pm 0.1)^\circ$ is almost absent, 
and thus we expect a non-zero $B_s$ decay width difference
\begin{equation}
    y_s \equiv \frac{\Gamma_{\rm L}-\Gamma_{\rm H}}{\Gamma_{\rm L}+\Gamma_{\rm H}}.
    %= \frac{\tau_{\rm H}-\tau_{\rm L}}{\tau_{\rm H}+\tau_{\rm L}},
\end{equation}
Indeed, two recent theoretical estimates of this quantity in the Standard Model are $y_s^{\rm SM} = 0.067 \pm 0.016$~\cite{Lenz:2011ti} and $y_s^{\rm SM} = 0.074 \pm 0.007$~\cite{Silvestrini:2013xx}.
These can be compared with the current most accurate determination of this quantity by the LHCb experiment: $y_s = 0.075 \pm 0.012$~\cite{Aaij:2013oba}, which confirms the presence of a non-zero decay width difference.

A non-zero width difference gives rise to interesting phenomenology for a non-flavour-tagged ({\it untagged}\/) sample of $B_s$ mesons.
Given such a sample, the time-dependent decay rate to a final state $f$ is the sum of the exponential decays of the two mass eigenstates\footnote{We assume no asymmetry in the number of produced $B_s^0$ and $\overline{B}_s^0$ mesons. 
A small production asymmetry can be neglected due to the rapid oscillation of $B_s$ mesons, which suppresses its effects by $\Gamma_s/\Delta M_s\approx 0.04$ at leading order.}~\cite{Dunietz:2000cr}
\begin{align}
   \langle \Gamma (B_s(t)\to f)\rangle
   &\equiv \Gamma (B^0_s(t)\to f) + \Gamma (\overline{B}^0_s(t)\to f) \notag\\
   &= \Gamma(B_{s,\rm H}\to f)\,e^{-\Gamma_{\rm H}\, t}
   + \Gamma(B_{s,\rm L}\to f)\,e^{-\Gamma_{\rm L}\, t} + {\cal O}(a)
   \label{untaggedRate}
\end{align}
where $a = \left|\Gamma_{12}/M_{12}\right|\sin\tilde\phi_s$ is vanishingly small for our purposes and can be neglected.
If the lifetimes of the two mass eigenstates differ, as we have observed they do, then a time-dependent analysis is sensitive not only to the total summed rate, $\Gamma(B_{s,\rm H}\to f)+ \Gamma(B_{s,\rm L}\to f)$, but also to the mass eigenstate rate asymmetry:
\begin{equation}
    {\cal A}_{\Delta\Gamma}^f = \frac{\Gamma(B_{s,\rm H}\to f)- \Gamma(B_{s,\rm L}\to f)}{\Gamma(B_{s,\rm H}\to f)+ \Gamma(B_{s,\rm L}\to f)}.
\end{equation}
As we will see, this final state dependent observable, accessible only if $y_s\neq 0$, is potentially sensitive to New Physics (NP).  

The layout of this write up is as follows. 
We first discuss the effect of a non-zero width difference on the definition of a $B_s$ branching ratio. 
Next we investigate how effective lifetime measurements, which probe ${\cal A}_{\Delta\Gamma}^f$, can constrain the $B_s^0$--$\overline{B}_s^0$ mixing parameters.
Our attention then shifts to the rare decay $B_s\to \mu^+ \mu^-$, for which we present 
the new observable ${\cal A}_{\Delta\Gamma}^{\mu\mu}$ that is accessible from a time-dependent analysis of this decay.
Finally, we consider how the combination of the $B_s\to \mu^+ \mu^-$ branching ratio and ${\cal A}_{\Delta\Gamma}^{\mu\mu}$ can help to distinguish between models of NP.

% #############################################################################
\section{Defining a $B_s$ branching ratio\label{sec:BR}}

At hadron colliders a $B_s$ branching ratio to a specific final state $f$ is measured by counting all untagged events over the accessible decay time, and normalizing this count against a known branching ratio measured at a B-factory.
It is therefore defined with respect to the untagged time-dependent rate given in \eqref{untaggedRate} as~\cite{Dunietz:2000cr,DeBruyn:2012wj} 
\begin{equation}
    \overline{\rm BR}(B_s \to f) \equiv \frac{1}{2} \int_0^\infty   
    \langle \Gamma (B_s(t)\to f)\rangle\, dt 
    = \frac{1}{2}\left[\frac{\Gamma(B_{s,\rm H}\to f)}{\Gamma_{\rm H}} + \frac{\Gamma(B_{s,\rm L}\to f)}{\Gamma_{\rm L}} \right].
    \label{BRbarDefn}
\end{equation}
This definition gives the average branching ratio of the two mass eigenstates.

From a theoretical perspective, squared amplitudes for $B_q\to f$ transitions ($q=s,d$) are customarily computed in the flavour basis rather than the mass eigenstate basis. 
%\begin{equation}
%    \Gamma(B_s^0\to f) + \Gamma(\overline{B}_s^0\to f) 
%    = \left.\langle \Gamma(B_s(t)\to f)\rangle\right|_{t=0}.
%\end{equation}
However, the total decay width (or lifetime) of a non-mass eigenstate is not well defined,
so instead the average $B_q$ decay width $(\Gamma_{\rm H} + \Gamma_{\rm L})/2 = 1/\tau_{B_q}$ is substituted.
This leads to the following definition for a $B_q$ branching ratio:
\begin{align}
    {\rm BR}(B_q \to f) \equiv  \frac{\tau_{B_q}}{2} \left.\langle \Gamma(B_q(t)\to f)\rangle\right|_{t=0}
    &= \frac{1}{2}\left[\frac{\Gamma(B^0_{q}(t)\to f)\big|_{t=0}}{\frac{1}{2}(\Gamma_{\rm H}+\Gamma_{\rm L})} + \frac{\Gamma(\overline{B}^0_{q}(t)\to f)\big|_{t=0}}{\frac{1}{2}(\Gamma_{\rm H}+\Gamma_{\rm L})} \right] \notag\\
    &= \frac{1}{2}\left[\frac{\Gamma(B_{q,\rm H}\to f)}{\frac{1}{2}(\Gamma_{\rm H}+\Gamma_{\rm L})} + \frac{\Gamma(B_{q,\rm L}\to f)}{\frac{1}{2}(\Gamma_{\rm H}+\Gamma_{\rm L})} \right],
    \label{BRDefn}
\end{align}
where the last equality can be directly compared to \eqref{BRbarDefn}.
We observe that if $\Gamma_{\rm H}=\Gamma_{\rm L}$ ($y_s=0$) the definitions in \eqref{BRbarDefn} and \eqref{BRDefn} are equivalent.
This is effectively true for $B_d$ mesons, where $y_d\approx 0$.
However, having established that $y_s\neq 0$, it becomes necessary to be able to convert between the theoretical and experimental definitions for the $B_s$ meson system.
The dictionary to convert between these definitions is compactly given by~\cite{DeBruyn:2012wj}
\begin{equation}
     \overline{\rm BR}(B_s \to f) = {\rm BR}(B_s \to f)
     \left[\frac{1 + y_s\,{\cal A}_{\Delta\Gamma}^f}{1-y_s^2}\right].
     \label{BRdict}
\end{equation}
Thus the required correction depends on the final state specific quantity ${\cal A}_{\Delta\Gamma}^f$.
This quantity can either be extracted from a time-dependent measurement (such as an effective lifetime as we discuss in the next section) or computed theoretically.

% #############################################################################
\section{$B^0_s$-$\overline{B}^0_s$ mixing constraints from effective lifetimes}

The observable ${\cal A}_{\Delta\Gamma}^f$ appears with the following functional form in the time dependent untagged rate given in \eqref{untaggedRate}:
\begin{align}
    \langle \Gamma(B_q(t)\to f)\rangle
    = &\left[\Gamma(B_{q,\rm H}\to f) + \Gamma(B_{q,\rm L}\to f) \right] \notag\\
    &\ \times\ e^{-t/\tau_{B_s}}\left\{ \cosh\left(y_s\, t/ \tau_{B_s}\right)
        + \sinh\left(y_s\, t/ \tau_{B_s}\right)\,{\cal A}_{\Delta\Gamma}^f \right\}.
\end{align}
It can be extracted, along with $y_s$, by fitting this function to time-dependent data. 
Alternatively, given low statistics, experiments also fit a single exponential to \eqref{untaggedRate}, which gives the {\it effective lifetime} $\tau^f_{\rm eff}$.
A maximum likelihood fit gives~\cite{DeBruyn:2012wj}
\begin{equation}
    \tau^f_{\rm eff} = \frac{\int^\infty_0 t\, \langle \Gamma(B_q(t)\to f)\rangle\,dt}
    {\int^\infty_0 \langle \Gamma(B_q(t)\to f)\rangle\,dt}
    = \frac{\tau_{B_s}}{1-y_s^2} \left(\frac{1+2\,y_s\, {\cal A}_{\Delta\Gamma}^f+ y_s^2}{1+ y_s\, {\cal A}_{\Delta\Gamma}^f}\right).
    \label{effLifetime}
\end{equation}
Thus the effective lifetime is a probe of the mass eigenstate rate asymmetry ${\cal A}_{\Delta\Gamma}^f$.
An interesting application of the effective lifetime is that it can be substituted into the branching ratio dictionary \eqref{BRdict} to give~\cite{DeBruyn:2012wj}
\begin{equation}
     {\rm BR}(B_s \to f) = \overline{\rm BR}(B_s \to f)
     \left[2 - (1-y_s^2)\frac{\tau_{\rm eff}^f}{\tau_{B_s}}\right].
\end{equation}
In this form the theoretical branching ratio is expressed entirely in terms of measurable quantities.

Now consider a CP eigenstate final state $f$ such that $ CP|f\rangle = \eta_f |f\rangle$ with CP eigenvalue $\eta_f=\pm 1$.
Such a final state is subject to mixing-induced CP violation, so that
\begin{equation}
    {\cal A}_{\Delta\Gamma}^f = -\eta_f \sqrt{1-C_f} \cos(\phi_s + \Delta\phi_f),
     \label{ADGCpState}
\end{equation}
where $\phi_s$ is the $B^0_s$-$\overline{B}_s^0$ mixing phase, 
$C_f$ parameterises direct CP violation in the decay mode and $\Delta\phi_f$ is a phase shift induced by the decay mode.
% and
%\begin{align}
%    C_f = \frac{1 - \left|A(\overline{B}_s^0\to f)/A({B}_s^0\to f)\right|^2}{1 + \left|A(\overline{B}_s^0\to f)/A({B}_s^0\to f)\right|^2},\qquad
%    \Delta\phi_f = -{\rm arg}\left(A(\overline{B}_s^0\to f)/A({B}_s^0\to f) \right)
%\end{align}
%are the direct CP violation and hadronic phase shift of the decay mode respectively.
In the absence of CP violating phases in the mixing and decay we have ${\cal A}_{\Delta\Gamma}^f = \pm 1$, implying that the decay proceeds solely via the heavy or light mass eigenstate as expected.
A measurement different from $\pm 1$ is thereby indicative of CP violating phases and potentially New Physics contributions.

By substituting \eqref{ADGCpState} into \eqref{effLifetime} we observe that an effective lifetime measurement constrains the $B_s$ mixing parameters $y_s$ and $\phi_s$, provided we know the decay mode specific parameters $C_f$ and $\Delta\phi_f$.
If we assume for simplicity that $C_f=\Delta\phi_f=0$, we find two possible solutions: ${\cal A}_{\Delta\Gamma}^f=\mp \cos\phi_s$.
The effective lifetime measurements for a CP even or odd decay mode will in principle result in two distinct contours in the $\phi_s$--$y_s$ plane,
with their intersection pinpointing the solution.

The LHCb experiment has currently measured effective lifetimes for a number of $B_s$ final states:
the CP even $K^+K^-$~\cite{Aaij:2012ns} and the CP odd $J/\psi\, f_0(980)$~\cite{Aaij:2012nta}, $J/\psi\, \pi^+\pi^-$~\cite{Aaij:2013oba} and $J/\psi\, K_S$~\cite{Playfer:2013xx}.
Although the $J/\psi\, \pi^+\pi^-$ effective lifetime measurement is more accurate than $J/\psi\, f_0(980)$, the two pions are produced by a sum of intermediate states that make the hadronic physics in this decay mode harder to control.
We also exclude the $J/\psi\, K_S$ effective lifetime on account of its larger errors.
This leaves the pair of CP even and odd final states $B_s\to K^+ K^-$ and $B_s\to J/\psi\,f_0(980)$, for which we proceed to control the decay mode parameters $C_f$ and $\Delta\phi_f$.

The CP asymmetries of the CP even mode $B_s\to K^+ K^-$ have been measured~\cite{LHCb-CONF-2012-007}, but the errors are still large.
A more accurate control of this decay mode is possible by employing its $SU(3)$ flavour symmetry relations to $B_d \to \pi^+ \pi^-$ ($U$-spin) and $B_d \to \pi^\mp K^\pm$~\cite{Fleischer:1999pa,Fleischer:2010ib}.
In Ref.~\cite{Fleischer:2010ib} such an analysis is performed, giving $C_{K^+ K^- } = 0.09\pm 0.04$ and $\Delta\phi_{K^+ K^-} = -\left(10.5^{+3.1}_{-2.8}\right)^\circ$, as well as the CKM angle $\gamma = (68\pm 7)^\circ$, which is in good agreement with fits of the CKM unitarity triangle.
%~\cite{UTfit,CKMfitter}.

An important issue for controlling the direct CP violation and hadronic phase shift in the CP odd mode $B_s\to J/\psi\,f_0(980)$ is the uncertain nature of the scalar $f_0(980)$ state.
At its simplest it is a pure quark-anti-quark $s\bar s$ state, however, there may be a mixing contribution from other isospin singlet states.
Alternatively, a popular interpretation of this scalar state, which explains better its light mass, is that of a tetraquark.
The picture we choose to subscribe to affects the contributing decay topologies, and in turn the magnitude of the decay mode parameters.
Fortunately, all conceivable topologies with CP violating phases are doubly Cabibbo suppressed~\cite{Fleischer:2011au,Fleischer:2011ib}.
Assuming no hierarchy between hadronic decay topologies, we find $\Delta\phi_{J/\psi f_0} \in [-3^\circ, 3^\circ]$ and $C_{J/\psi f_0} \lesssim 0.05$~\cite{Fleischer:2011au}.
%The situation is similar for the CP even final state modes $B_s\to J/\psi \eta^{(\prime)}$~\cite{Fleischer:2011ib}, for which we await effective lifetime measurements.

\begin{figure}
\begin{center}
\includegraphics[height=4cm]{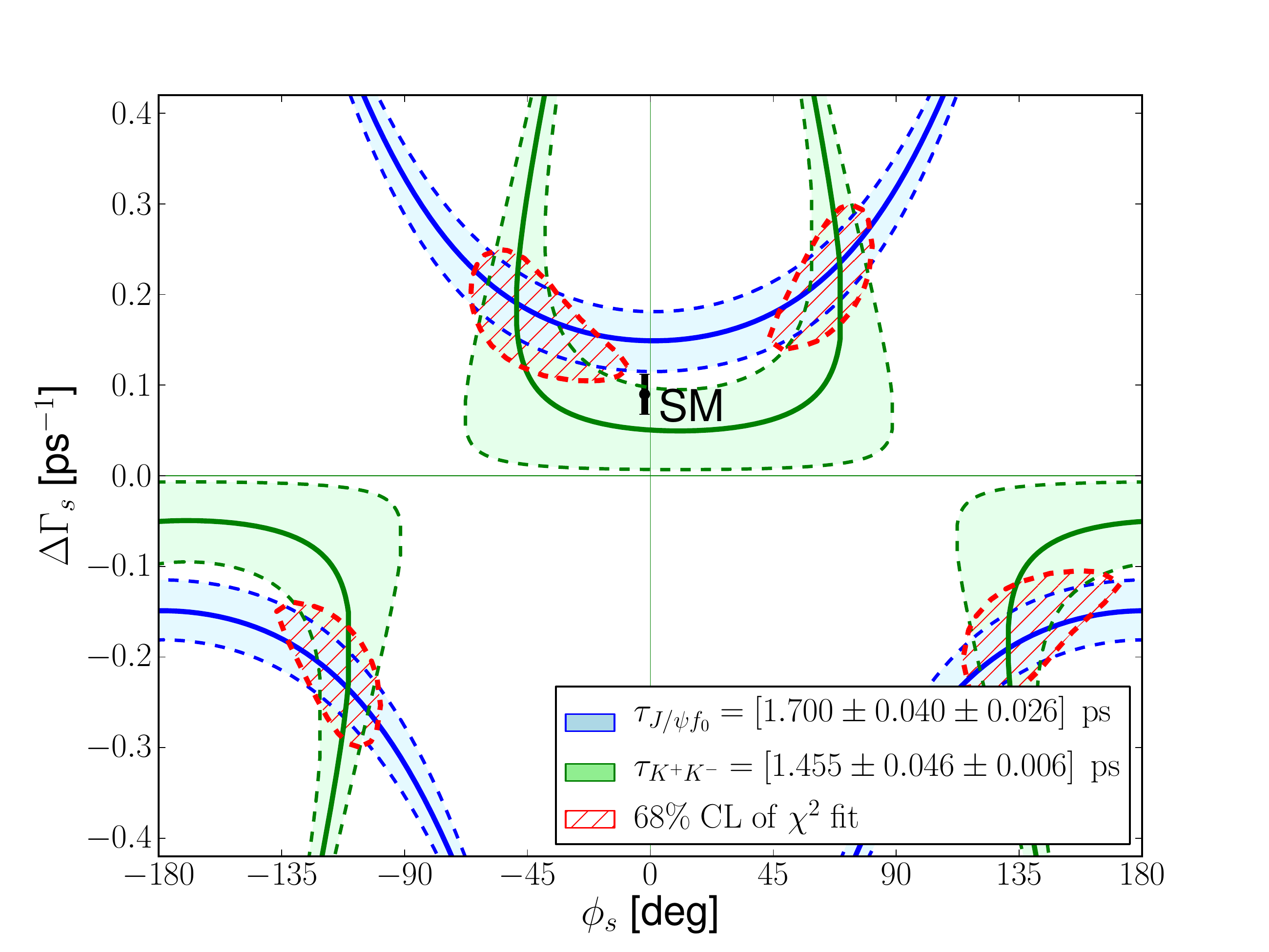}
\includegraphics[height=4cm]{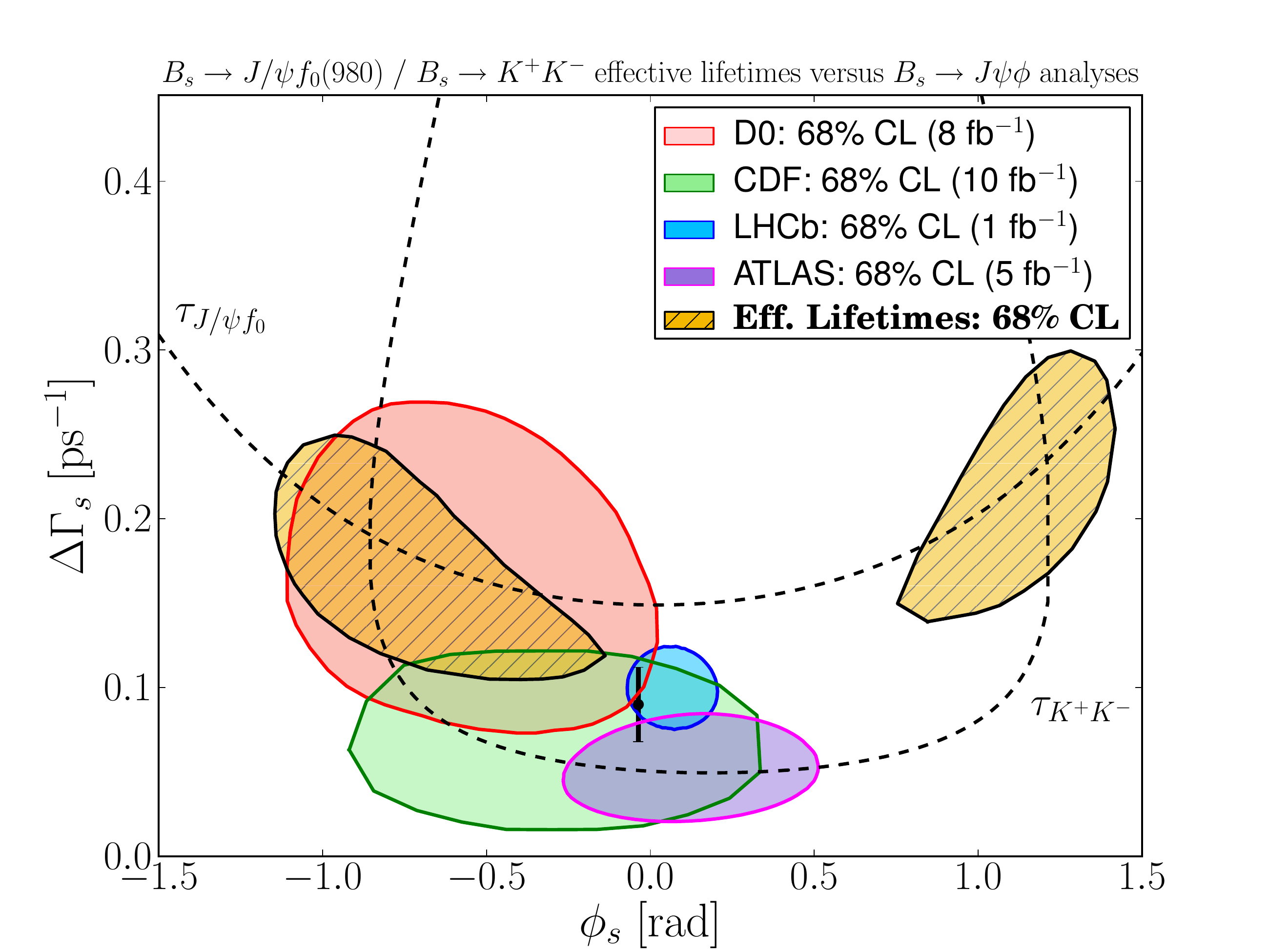}
\end{center}
\caption{The $B_s\to K^+ K^-$ and $B_s\to J/\psi f_0(980)$ effective lifetimes  projected onto the $\phi_s$--$\Delta\Gamma_s$ ($y_s$) plane. {\it Left panel:} shaded bands show 1\,$\sigma$ errors of current measurements and resulting 68\% CL $\chi^2$ fit. {\it Right panel:} Comparison with fits from full tagged analyses of the $B_s\to J/\psi\phi$ decay mode from several experiments.}
\label{fig:DGphiS}
\end{figure}

We now have the ingredients necessary to plot our pair of CP even and odd effective lifetimes in the $\phi_s$--$\Delta\Gamma_s$ plane (where $\Delta\Gamma_s = 2\,y_s/\tau_{B_s}$).
In the left panel of Figure~\ref{fig:DGphiS} we show these contours, as well as the 68 \% CL regions corresponding to their intersection. 
These regions currently lie 1\,$\sigma$ from the SM prediction.
In the right panel of Figure~\ref{fig:DGphiS} we compare our effective lifetime analysis with the full tagged $B_s\to J/\psi \phi$ analyses of LHCb, ATLAS, D\O\ and CDF experiments.
% ATLAS ~\cite{Heller:2013xx}
This comparison shows that effective lifetimes give a complementary determination of these mixing parameters.
%However, they struggle to compete with a full tagged analysis, especially for small $\phi_s$.
%Untagged time-dependent analyses may therefore be most useful for $B_s$ decays where a tagged analysis lies far in future, an example of which we discuss in the next two sections.

% #############################################################################
\section{Time-dependent analysis of $B_s\to\mu^+\mu^-$}

The decay $B_s\to \mu^+\mu^-$ is famous among its peers for being both very suppressed in the SM, and very sensitive to NP~\cite{Altmann:2013xx}.
Indeed, it was long hoped that NP would induce a branching ratio orders of magnitude larger than its SM prediction.
The LHCb experiment has now measured this branching ratio to be $\overline{\rm BR}(B_s\to\mu^+\mu^-)_{\rm LHCb} = (3.2^{+1.5}_{-1.2})\times 10^{-9}$~\cite{Aaij:2012nna}.
To compare this number with the SM prediction we need to use the correct branching ratio definition.
As we will soon discuss in more detail, in the Standard Model only the heavy mass eigenstate contributes to the two muon final state~\cite{DeBruyn:2012wk}.
Thus we should use ${\cal A}^{\mu\mu}_{\Delta\Gamma,\rm SM}=1$ in our dictionary \eqref{BRdict}, and we consequently find~\cite{Buras:2013uqa}
\begin{equation}
    \overline{\rm BR}(B_s\to\mu^+\mu^-)_{\rm SM}
    = (3.56 \pm 0.18)\times 10^{-9}.
\end{equation}
The room left for NP to alter this total rate is thus moderate at best, but, as we will argue, this may not be our only available probe.
For later convenience we define the parameter
\begin{equation}
    \overline{R} \equiv \frac{\overline{\rm BR}(B_s\to\mu^+\mu^-)}{\overline{\rm BR}(B_s\to\mu^+\mu^-)_{\rm SM}},
\end{equation}
so that $\overline{R}_{\rm LHCb}=0.90^{+0.42}_{-0.34}$ and $\overline{R}_{\rm SM}=1$.

\subsection{Time-dependent observables}

It is useful to discuss NP contribution to $B_s\to\mu^+\mu^-$ in the language of an effective field theory~\cite{Altmannshofer:2011gn}:
\begin{equation}
	{\cal H}_{\rm eff} = -\frac{G_F\,\alpha}{\sqrt{2}\pi}\left\{
		V_{ts}^* V_{tb}\,  \sum_{i}^{10,S,P} \left( C_i\,{\cal O}_i + C'_i\,{\cal O}'_i\right) 
		+ {\rm h.c}\right\},
		\label{effOPE}
\end{equation}
with
${\cal O}_{10} = (\bar s\gamma_\mu P_L b)(\bar l \gamma^\mu \gamma_5 l)$, 
${\cal O}_{S} = m_b (\bar s P_R b)(\bar l l),$ and
${\cal O}_{P} = m_b (\bar s P_R b)(\bar l \gamma_5 l)$, 
%\begin{align}
%	{\cal O}_{10} = (\bar s\gamma_\mu P_L b)(\bar l \gamma^\mu \gamma_5 l), \quad
%	{\cal O}_{S} = m_b (\bar s P_R b)(\bar l l),\quad
%	{\cal O}_{P} = m_b (\bar s P_R b)(\bar l \gamma_5 l), 
%\end{align}
and the primed operators given by the interchange $P_L\leftrightarrow P_R$.
In the SM only $C_{10}=C_{10}^{\rm SM}$ is not vanishingly small.
It turns out that our observables of interest can be expressed purely in terms of the parameters~\cite{DeBruyn:2012wk}
\begin{align}
	P &\equiv \frac{C_{10} - C'_{10}}{C_{10}^{\rm SM}} + \frac{m_{B_s}^2}{2m_\mu}\left(\frac{m_b}{m_b + m_s}\right) \left( \frac{C_{P} - C'_{P}}{C_{10}^{\rm SM}}\right)\equiv |P|e^{i\varphi_P}, \notag\\
	S &\equiv \sqrt{1 - \frac{4\,m_\mu^2}{m_{B_s}^2}}\frac{m_{B_s}^2}{2m_\mu}\left(\frac{m_b}{m_b + m_s}\right) \left( \frac{C_{S} - C'_{S}}{C_{10}^{\rm SM}}\right)\equiv |S|e^{i\varphi_S}.
	\label{PSdefn}
\end{align}
Specifically, the mass eigenstate decay rates depend on $P$ and $S$ as
\begin{align}
    \Gamma(B_{s,\rm H}\to \mu^+\mu^-) &\propto |P|^2 \cos^2(\varphi_P - \phi_s^{\rm NP}/2) + |S|^2\sin^2(\varphi_S - \phi_s^{\rm NP}/2), \notag\\    
    \Gamma(B_{s,\rm L}\to \mu^+\mu^-) &\propto |P|^2 \sin^2(\varphi_P - \phi_s^{\rm NP}/2) + |S|^2\cos^2(\varphi_S - \phi_s^{\rm NP}/2),
\end{align}
where $\phi_s^{\rm NP} \equiv \phi_s - \phi_s^{\rm SM}$ parameterises possible NP in $B_s$ mixing.
We see that in the SM, where $P= 1$, $S= 0$ and $\phi_s^{\rm NP}= 0$, only the heavy mass eigenstate contributes, and ${\cal A}^{\mu\mu}_{\Delta\Gamma,\rm SM}=1$.
In general we have~\cite{DeBruyn:2012wk}
\begin{align}
    \overline{R} = \left[\frac{1 + y_s\,{\cal A}^{\mu\mu}_{\Delta\Gamma}}{1 + y_s}\right]\left(|P|^2 + |S|^2 \right),\quad  
    {\cal A}^{\mu\mu}_{\Delta\Gamma} = \frac{|P|^2\cos(2\varphi_P-\phi_s^{\rm NP}) - |S|^2\cos(2\varphi_S-\phi_s^{\rm NP})}{|P|^2 + |S|^2}.\label{RADG}
\end{align}
Thus an untagged time-dependent analysis of $B_s\to \mu^+\mu^-$, such as an effective lifetime measurement, would give us the independent and theoretically clean new observable ${\cal A}^{\mu\mu}_{\Delta\Gamma}$, complementing the branching ratio ($\overline{R}$).

We note in passing that if a tagged time-dependent measurement is possible, a third observable, ${\cal S}_{\mu\mu}$, is accessible from the flavour state rate asymmetry.
This observable is related to ${\cal A}^{\mu\mu}_{\Delta\Gamma}$ by
\begin{equation}
    |{\cal A}^{\mu\mu}_{\Delta\Gamma}|^2 + |{\cal S}_{\mu\mu}|^2 = 1 - \left[\frac{2|P||S|\cos(\varphi_P - \varphi_S)}{|P|^2 + |S|^2} \right]^2.
\end{equation}
It is an independent observable if scalar operators are not negligible.

\subsection{Solvable scenarios of New Physics}

\begin{figure}
\begin{center}
\includegraphics[height=4cm]{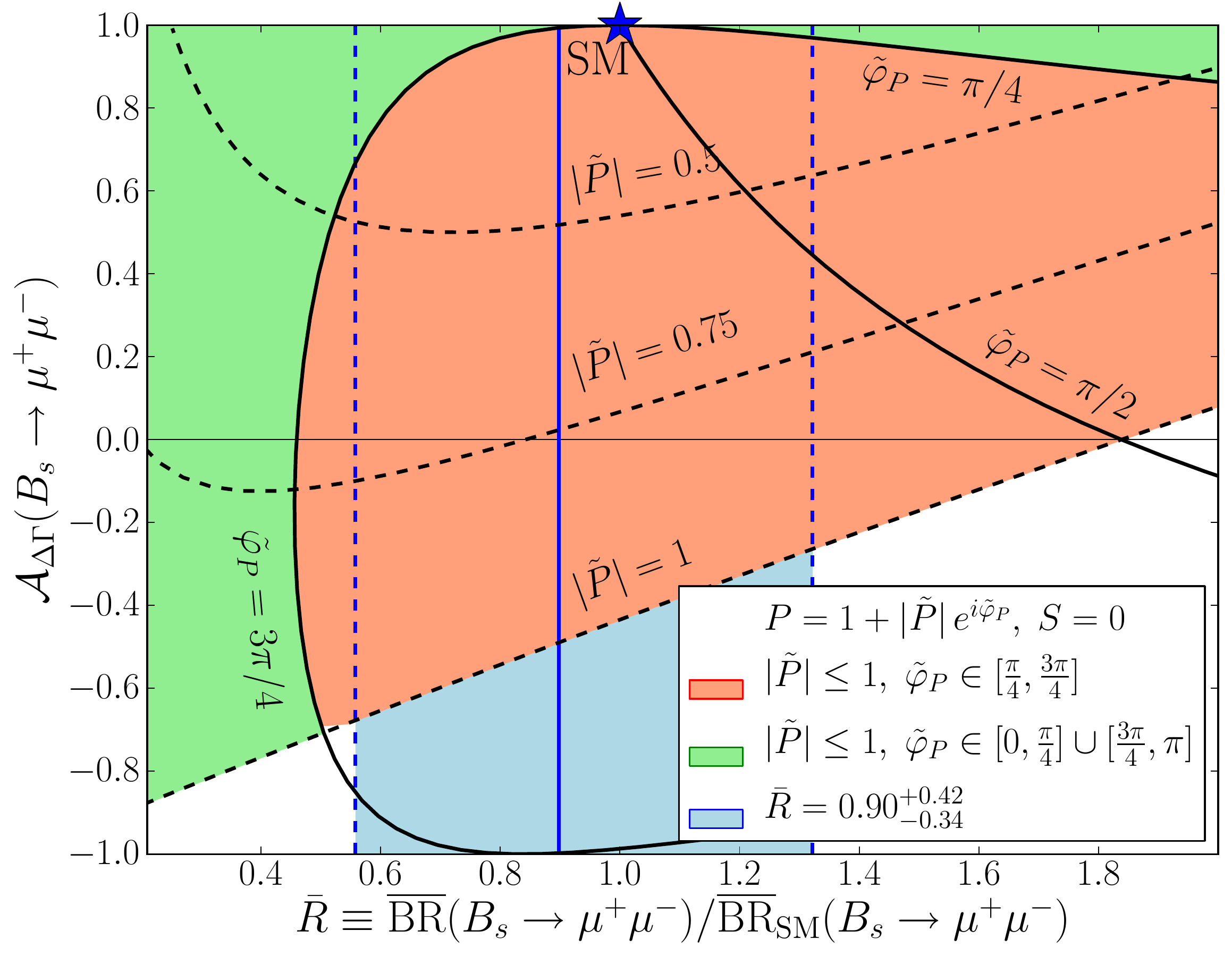}
\includegraphics[height=4cm]{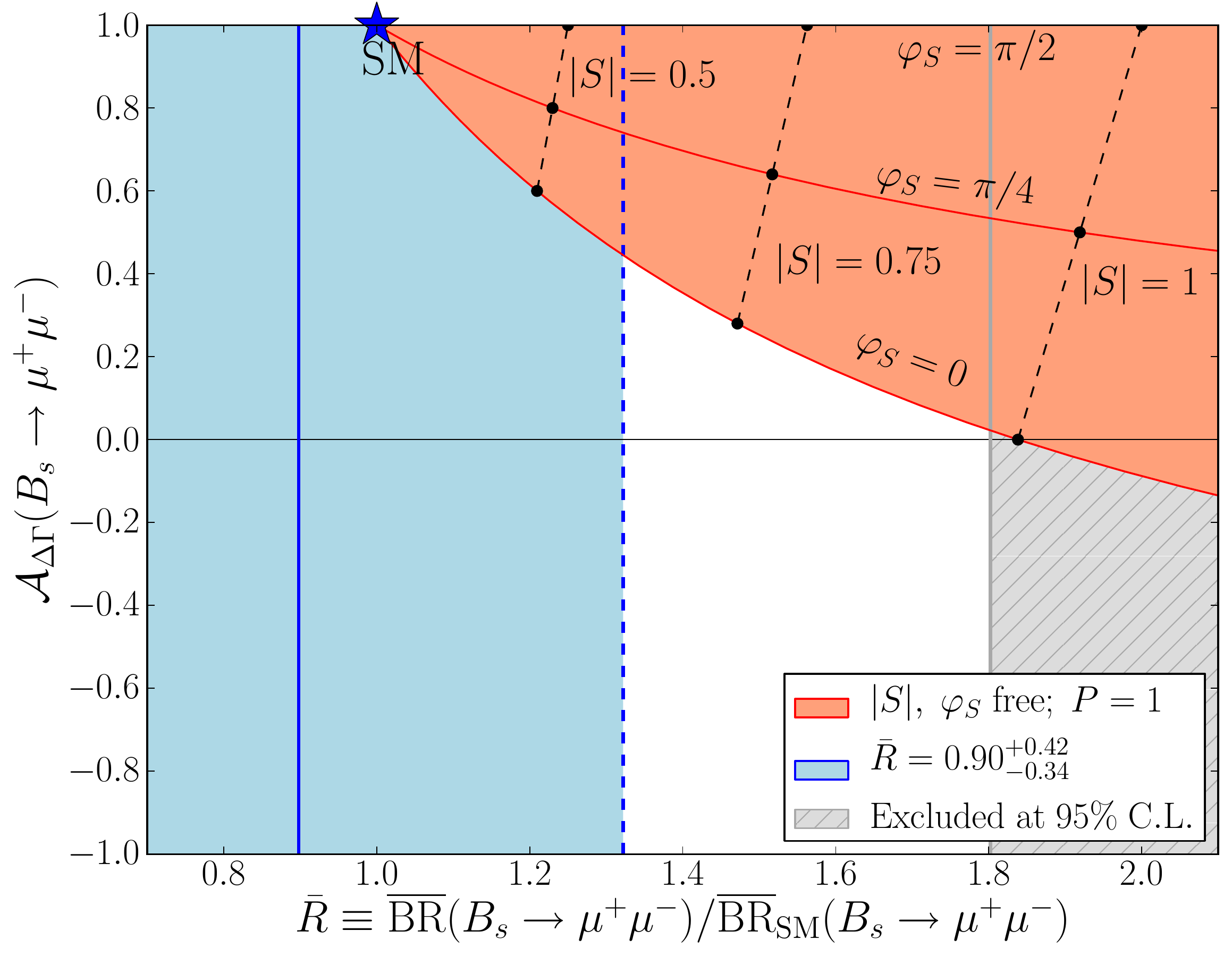}
\end{center}
\caption{Correlations between the $\overline{R}$ and ${\cal A}^{\mu\mu}_{\Delta\Gamma}$ observables in Scenario A (left panel) and Scenario B (right panel), as discussed in the text.}
\label{fig:R-ADG_scenAB}
\end{figure}

\begin{figure}
\begin{center}
\includegraphics[height=4cm]{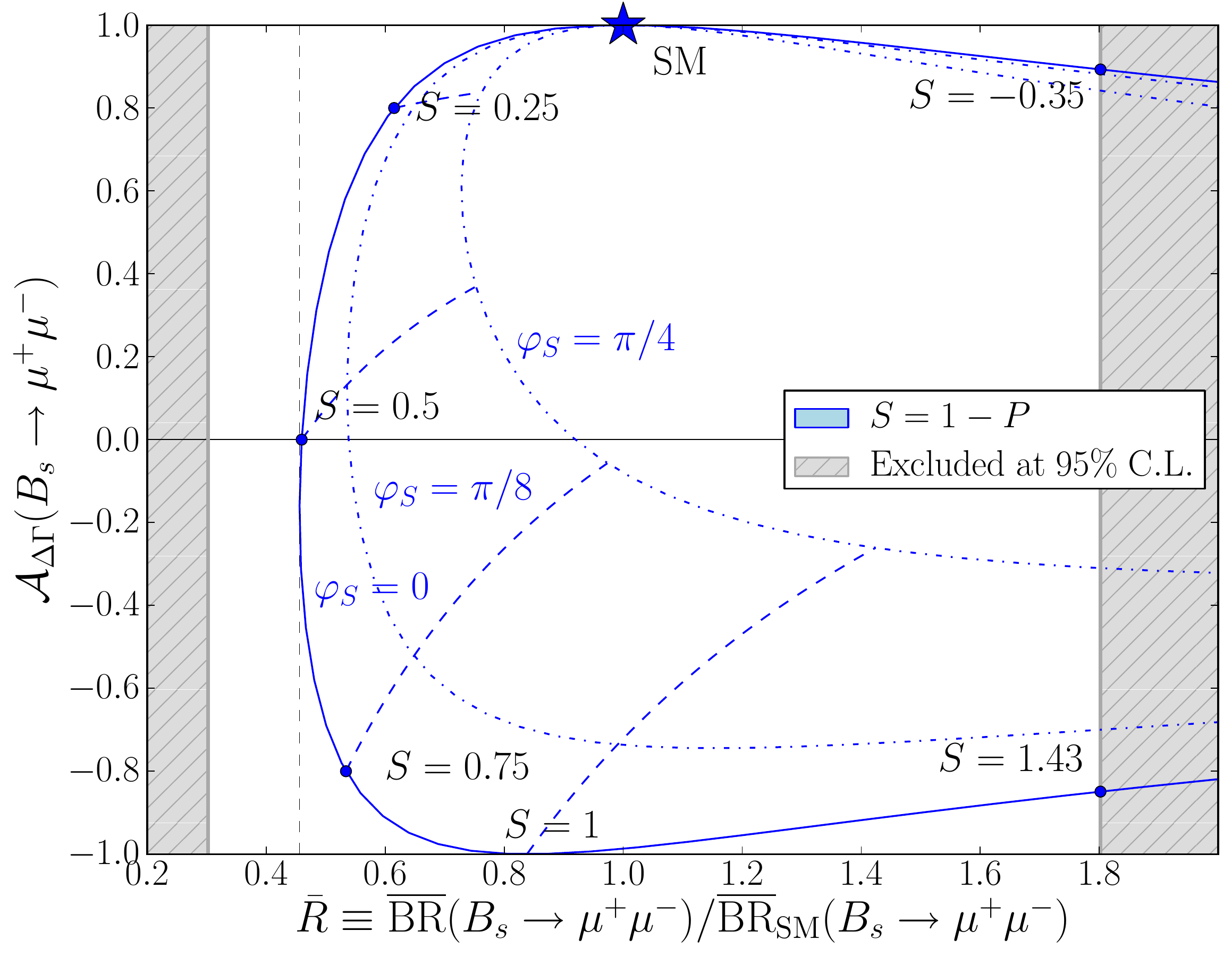}
\includegraphics[height=4.35cm]{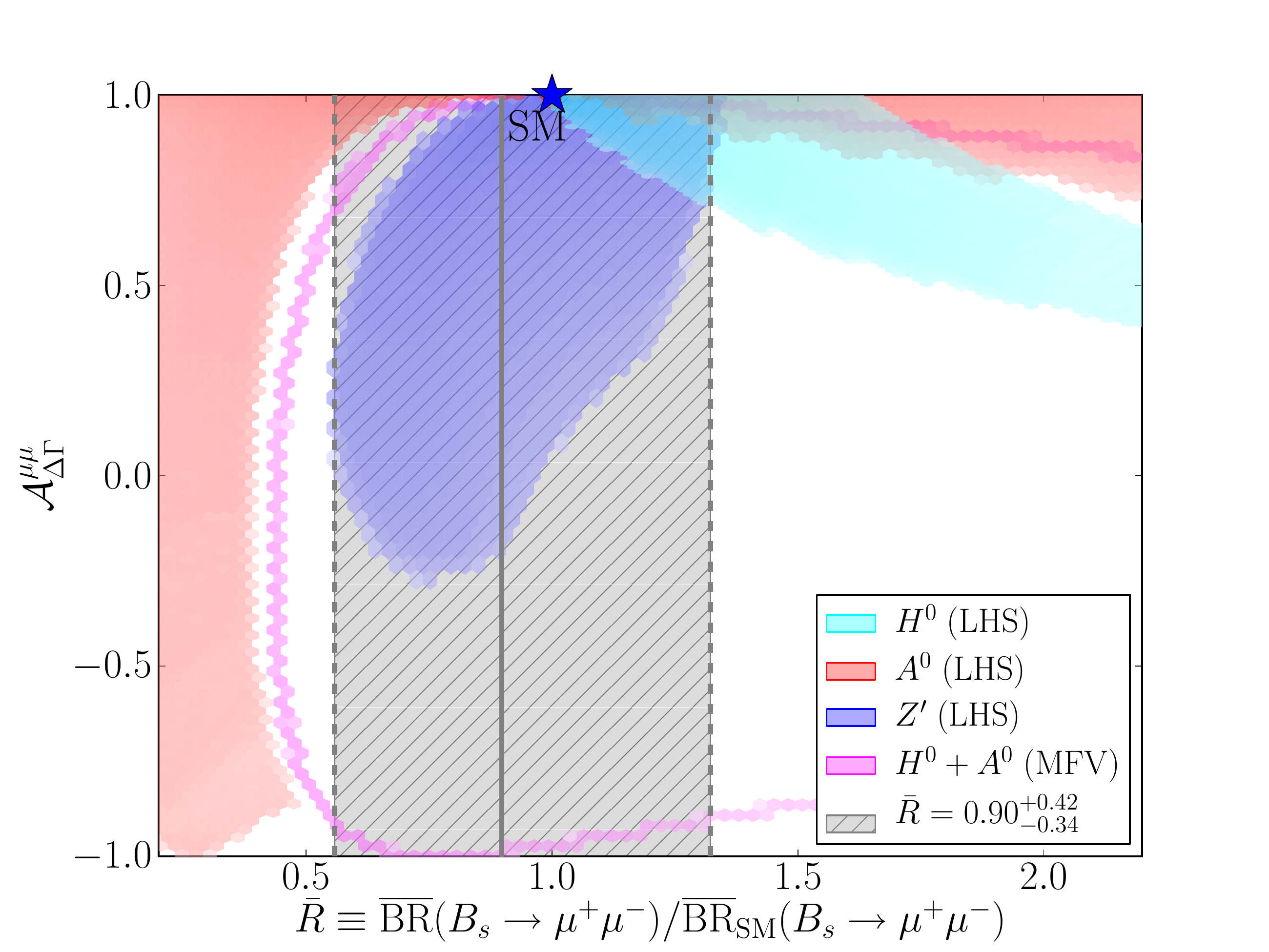}
\end{center}
\caption{Correlations between the $\overline{R}$ and ${\cal A}^{\mu\mu}_{\Delta\Gamma}$ observables. {\it Left panel:} Scenario C, as discussed in text. {\it Right panel:} experimentally allowed parameter space for specific models, with a mass of 1\,TeV for the mediating particle.}
\label{fig:R-ADG_scenC_grand}
\end{figure}

An untagged time-dependent analysis of $B_s\to \mu^+\mu^-$ would give us two independent observables depending in principle on four model-independent parameters: $|P|$, $|S|$, $\varphi_P$, $\varphi_S$.
We exclude $\phi_s^{\rm NP}$ from this counting as it should be well measured by the time a time-dependent analysis is feasible.
Thus in general we do not have sufficient observables to solve the system without invoking observables from other processes.
Yet there are a number of phenomenologically well-motivated scenarios where two observables are sufficient.
These scenarios are captured by the following constraints~\cite{Buras:2013uqa}:
\begin{itemize}
    \item {\bf Scenario A}: $S=0$. 
        In this scenario scalar operators either cancel or are not present to begin with. 
        The latter is realised in Constrained Minimal Flavour Violation, models with $Z'$ gauge bosons or two Higgs doublet models (2HDM) with a dominant pseudoscalar. 
        In the left panel of Figure~\ref{fig:R-ADG_scenAB} we sketch the parameter regions allowed in this scenario. 
        The CP violating phase $\varphi_P$ must differ from zero for ${\cal A}^{\mu\mu}_{\Delta\Gamma}$ to differ from one. 
        For moderate NP effects a sizable phase is needed to stay in the ballpark of the SM branching ratio.
    \item {\bf Scenario B}: $P=1$. 
        In this scenario NP effects are driven purely by new scalar operators. 
        For example a dominant heavy scalar particle in a 2HDM.
        In the right panel of Figure~\ref{fig:R-ADG_scenAB} we sketch the parameter regions allowed in this scenario. 
        Observe that the branching ratio can only increase. 
        For moderate NP the observable ${\cal A}^{\mu\mu}_{\Delta\Gamma}$ tends towards zero in the absence of a CP violating phase $\varphi_S$, 
        but is bounded to be positive by the branching ratio measurement.
    \item {\bf Scenario C}: $S=\pm(1-P)$. 
        Letting $P=1+\tilde{P}$ gives $S=\pm \tilde{P}$ i.e.\ NP contributions to $S$ and $P$ are on the same footing.
        An interesting example is decoupled 2HDMs, where the mass of the heavy scalar and pseudoscalar are much larger than that of the light scalar, so that $C^{(\prime)}_S= \mp C^{(\prime)}_P$.
        In the left panel of Figure~\ref{fig:R-ADG_scenC_grand} the solid blue line sketches this constraint in the absence of a CP violating phase.
        The effect of a CP violating phase is shown by the dashed lines.
        In this scenario ${\cal A}^{\mu\mu}_{\Delta\Gamma}=-1$ is reachable with moderate NP without CP violating phases.
        Also characteristic of this scenario is the lower bound $\overline{R} \geq (1-y_s)/2$.
    \item {\bf Scenario D}: $\varphi_P = \varphi_S = \{0,\pi\}$. 
        This scenario assumes no new CP violating phases.
        Thus models with Minimal Flavour Violation naturally conform to this scenario.
        From the expressions in \eqref{RADG} we see that $\overline{R}$ gives a circle in the $|S|^2$--$|P|^2$ plane,
        whereas ${\cal A}^{\mu\mu}_{\Delta\Gamma}$ gives a line through the origin.
        Thus a measurement of both observables would cleanly disentangle these two model-independent parameters up to discrete ambiguities.
\end{itemize}

To explore the allowed parameter space of these scenarios we consider specific models with general $b\to s$ flavour changing neutral currents coupled to two muons via an intermediate heavy particle.
We take the exchanged intermediate particle to be a $Z'$ gauge boson, $H^0$ scalar, $A^0$ pseudoscalar or a $H^0+A^0$ combination, with a mass of 1\,TeV~\cite{Buras:2013uqa,Buras:2013rqa,Buras:2012jb}.
For each specific model we consider constraints from $B_s$ mixing by enforcing that $\Delta M_s$ stays within 5\% of its central experimental value and $\phi_s$ within 2\,$\sigma$ of its experimental measurement.
%In Ref.~\cite{Buras:2012fs} NLO corrections for these $\Delta F =2$ tree-level processes have been calculated.
For the $Z'$ model we also include constraints on the $C^{(\prime)}_{10}$ Wilson coefficients from other $b\to s$ transition processes such as $B\to K^* \mu^+ \mu^-$~\cite{Altmannshofer:2012az}.
The results for these models are shown in the right panel of Figure~\ref{fig:R-ADG_scenC_grand}, 
from which we observe that there is still plenty of room for NP within the currently available constraints.

% #############################################################################
\section{Summary}

The experimentally established non-zero decay width difference in the $B_s$ meson system gives us access to the mass eigenstate rate asymmetry  ${\cal A}^{f}_{\Delta\Gamma}$ for a $B_s\to f$ transition.
This observable can be extracted from the untagged time-dependent decay rate by either a full fit or a measurement of the effective decay lifetime.
It is not only a necessary ingredient in converting between different definitions of the $B_s$ branching ratio, but also a sensitive probe of NP.
For instance, a pair of effective lifetimes for CP even and odd final states can be used to constrain the NP effects present in $B_s$ mixing.
For the rare decay $B_s\to \mu^+\mu^-$, where a tagged analysis may never be feasible, the observable ${\cal A}^{\mu\mu}_{\Delta\Gamma}$ serves as a complementary constraint for NP.
Specifically, the combination of the $B_s\to \mu^+\mu^-$ and ${\cal A}^{\mu\mu}_{\Delta\Gamma}$ can distinguish between different scenarios of NP.

% #############################################################################
\acknowledgments

I would very much like to thank the organizers of the Beauty 2013 conference for their invitation, hospitality and organization.
I would further like to thank R.~Fleischer, A.~J.~Buras, J.~Girrbach, F.~De~Fazio and M.~Nagai, for pleasant collaborations.
I thank R. Fleischer for his comments on this manuscript.

\bibliographystyle{JHEP}

\bibliography{allrefs}

\providecommand{\href}[2]{#2}\begingroup\raggedright\begin{thebibliography}{10}

\bibitem{Lenz:2011ti}
A.~Lenz and U.~Nierste, {\it {Numerical Updates of Lifetimes and Mixing
  Parameters of B Mesons}},  \href{http://xxx.lanl.gov/abs/1102.4274}{{\tt
  arXiv:1102.4274}}.

\bibitem{Silvestrini:2013xx}
L.~Silvestrini, {\it {CP Violation Theory}},  {\em these proceedings}.

\bibitem{Aaij:2013oba}
{\bf LHCb} Collaboration, R.~Aaij et~al., {\it {Measurement of $CP$ violation
  and the $B_s^0$ meson decay width difference with $B_s^0 \to J/\psi K^+K^-$
  and $B_s^0\to J/\psi\pi^+\pi^-$ decays}},
  \href{http://xxx.lanl.gov/abs/1304.2600}{{\tt arXiv:1304.2600}}.

\bibitem{Dunietz:2000cr}
I.~Dunietz, R.~Fleischer, and U.~Nierste, {\it {In pursuit of new physics with
  $B_s$ decays}},  {\em Phys.Rev.} {\bf D63} (2001) 114015,
  [\href{http://xxx.lanl.gov/abs/hep-ph/0012219}{{\tt hep-ph/0012219}}].

\bibitem{DeBruyn:2012wj}
K.~De~Bruyn, R.~Fleischer, R.~Knegjens, P.~Koppenburg, M.~Merk, et~al., {\it
  {Branching Ratio Measurements of $B_s$ Decays}},  {\em Phys.Rev.} {\bf D86}
  (2012) 014027, [\href{http://xxx.lanl.gov/abs/1204.1735}{{\tt
  arXiv:1204.1735}}].

\bibitem{Aaij:2012ns}
{\bf LHCb} Collaboration, R.~Aaij et~al., {\it {Measurement of the effective
  $B_s^0 \rightarrow K^+ K^-$ lifetime}},  {\em Phys.Lett.} {\bf B716} (2012)
  393--400, [\href{http://xxx.lanl.gov/abs/1207.5993}{{\tt arXiv:1207.5993}}].

\bibitem{Aaij:2012nta}
{\bf LHCb} Collaboration, R.~Aaij et~al., {\it {Measurement of the $B_s$
  effective lifetime in the $J/\psi f_0(980)$ final state}},  {\em
  Phys.Rev.Lett.} {\bf 109} (2012) 152002,
  [\href{http://xxx.lanl.gov/abs/1207.0878}{{\tt arXiv:1207.0878}}].

\bibitem{Playfer:2013xx}
S.~Playfer, {\it {LHCb $\sin 2\beta$, $\Delta M_d$ and $B_d\to J/\psi
  K^{*0}$}},  {\em these proceedings}.

\bibitem{LHCb-CONF-2012-007}
{\bf LHCb} Collaboration, {\it {Measurement of time-dependent $CP$ violation in
  charmless two-body $B$ decays}},  {\em LHCb-CONF-2012-007} (Feb, 2012).

\bibitem{Fleischer:1999pa}
R.~Fleischer, {\it {New strategies to extract Beta and gamma from $B_d \to
  \pi^+ \pi^-$ and $B_s \to K^+ K^-$}},  {\em Phys.Lett.} {\bf B459} (1999)
  306--320, [\href{http://xxx.lanl.gov/abs/hep-ph/9903456}{{\tt
  hep-ph/9903456}}].

\bibitem{Fleischer:2010ib}
R.~Fleischer and R.~Knegjens, {\it {In Pursuit of New Physics with $B^0_s \to
  K^+K^-$}},  {\em Eur.Phys.J.} {\bf C71} (2011) 1532,
  [\href{http://xxx.lanl.gov/abs/1011.1096}{{\tt arXiv:1011.1096}}].

\bibitem{Fleischer:2011au}
R.~Fleischer, R.~Knegjens, and G.~Ricciardi, {\it {Anatomy of $B^0_{s,d} \to
  J/\psi f_0(980)$}},  {\em Eur.Phys.J.} {\bf C71} (2011) 1832,
  [\href{http://xxx.lanl.gov/abs/1109.1112}{{\tt arXiv:1109.1112}}].

\bibitem{Fleischer:2011ib}
R.~Fleischer, R.~Knegjens, and G.~Ricciardi, {\it {Exploring CP Violation and
  $\eta$-$\eta'$ Mixing with the $B^0_{s,d} \to J/\psi \eta^{(\prime)}$
  Systems}},  {\em Eur.Phys.J.} {\bf C71} (2011) 1798,
  [\href{http://xxx.lanl.gov/abs/1110.5490}{{\tt arXiv:1110.5490}}].

\bibitem{Altmann:2013xx}
W.~Altmannshofer, {\it {Theory Review of $B_{s,d} \to \mu^+\mu^-$ Decays }},
  {\em these proceedings}.

\bibitem{Aaij:2012nna}
{\bf LHCb} Collaboration, R.~Aaij et~al., {\it {First evidence for the decay
  $B_s \to \mu^+ \mu^-$}},  {\em Phys.Rev.Lett.} {\bf 110} (2013) 021801,
  [\href{http://xxx.lanl.gov/abs/1211.2674}{{\tt arXiv:1211.2674}}].

\bibitem{DeBruyn:2012wk}
K.~De~Bruyn, R.~Fleischer, R.~Knegjens, P.~Koppenburg, M.~Merk, et~al., {\it
  {Probing New Physics via the $B^0_s\to \mu^+\mu^-$ Effective Lifetime}},
  {\em Phys.Rev.Lett.} {\bf 109} (2012) 041801,
  [\href{http://xxx.lanl.gov/abs/1204.1737}{{\tt arXiv:1204.1737}}].

\bibitem{Buras:2013uqa}
A.~J. Buras, R.~Fleischer, J.~Girrbach, and R.~Knegjens, {\it {Probing New
  Physics with the $B_s \to {\mu}^+ {\mu}^-$ Time-Dependent Rate}},
  \href{http://xxx.lanl.gov/abs/1303.3820}{{\tt arXiv:1303.3820}}.

\bibitem{Altmannshofer:2011gn}
W.~Altmannshofer, P.~Paradisi, and D.~M. Straub, {\it {Model-Independent
  Constraints on New Physics in $b\to s$ Transitions}},  {\em JHEP} {\bf 1204}
  (2012) 008, [\href{http://xxx.lanl.gov/abs/1111.1257}{{\tt
  arXiv:1111.1257}}].

\bibitem{Buras:2013rqa}
A.~J. Buras, F.~De~Fazio, J.~Girrbach, R.~Knegjens, and M.~Nagai, {\it {The
  Anatomy of Neutral Scalars with FCNCs in the Flavour Precision Era}},
  \href{http://xxx.lanl.gov/abs/1303.3723}{{\tt arXiv:1303.3723}}.

\bibitem{Buras:2012jb}
A.~J. Buras, F.~De~Fazio, and J.~Girrbach, {\it {The Anatomy of Z' and Z with
  Flavour Changing Neutral Currents in the Flavour Precision Era}},  {\em JHEP}
  {\bf 1302} (2013) 116, [\href{http://xxx.lanl.gov/abs/1211.1896}{{\tt
  arXiv:1211.1896}}].

\bibitem{Altmannshofer:2012az}
W.~Altmannshofer and D.~M. Straub, {\it {Cornering New Physics in $b\to s$
  Transitions}},  {\em JHEP} {\bf 1208} (2012) 121,
  [\href{http://xxx.lanl.gov/abs/1206.0273}{{\tt arXiv:1206.0273}}].

\end{thebibliography}\endgroup

%\begin{thebibliography}{99}
%\end{thebibliography}

\end{document}